\documentclass[a4paper]{article}

\usepackage{INTERSPEECH2018, amsmath, multirow, epstopdf}

\title{Investigating the role of L1 in automatic pronunciation evaluation of L2 speech}
\name{Ming Tu$^1$, Anna Grabek$^2$, Julie Liss$^1$, Visar Berisha$^{1,2}$}
\address{
  $^1$Speech and Hearing Science Department\\
  $^2$School of Electrical, Computer, and Energy Engineering \\
  Arizona State University}
\email{\{mingtu, agrabek, jmliss, visar\}@asu.edu}

\begin{document}

\maketitle
\begin{abstract}
Automatic pronunciation evaluation plays an important role in pronunciation training and second language education. This field draws heavily on concepts from automatic speech recognition (ASR) to quantify how close the pronunciation of non-native speech is to native-like pronunciation. However, it is known that the formation of accent is related to pronunciation patterns of both the target language (L2) and the speaker's first language (L1). In this paper, we propose to use two native speech acoustic models, one trained on L2 speech and the other trained on L1 speech. We develop two sets of measurements that can be extracted from two acoustic models given accented speech. A new utterance-level feature extraction scheme is used to convert these measurements into a fixed-dimension vector which is used as an input to a statistical model to predict the accentedness of a speaker. On a data set consisting of speakers from 4 different L1 backgrounds, we show that the proposed system yields improved correlation with human evaluators compared to systems only using the L2 acoustic model.

\end{abstract}

\vspace{0.2cm}

\section{Introduction}

With the development of speech technologies, pronunciation training in second language education can be replaced by computer-based systems. Automatic pronunciation evaluation has always been an important part of Computer Assisted Pronunciation Training (CAPT). The goal of automatic pronunciation evaluation is to build an automatic system which can measure the quality of pronunciation given input speech. Automatic speech recognition (ASR) models play an important role in this area. The acoustic model in an ASR system trained on native speech provides a baseline distribution for each phoneme/word; new speech samples can be projected on this distribution to determine how statistically close the pronunciation is to a native pronunciation.

 From a speech learning perspective, accented speech is the result of second language (L2) speech being produced by a sensorimotor control system that has overlearned first language (L1) sound contrasts and rhythmic composition. The Speech Learning Model (SLM) plays an important role in explaining L2 speech learning, which is based on the idea that phonetic systems respond to L2 sounds by adding new phonetic categories, or by modifying existing L1 phonetic categories \cite{flege1995second}. The SLM emphasizes the interplay between L1 and L2 in forming the target language phonetic systems of language learners. Based on the SLP hypotheses, an equivalence classification is applied to an L2 phone similar to a previously experienced L1 category, thereby degrading the accuracy of L2 sound production. Since certain phonetic and phonological patterns can be transferred from L1 to the learned L2, English spoken by people from different L1 backgrounds show acoustic characteristics similar to the speakers' mother language \cite{chang2010first}\cite{jiao2016accent}.

However, almost all existing pronunciation evaluation systems only use acoustic models trained on native L2 (i.e. the target language) speech to extract useful measurements to quantify how close non-native speech is to the native pronunciation of the target language. For example, the study in \cite{franco1997automatic} proposed measurements based on both phoneme-level log-likelihood and posterior probabilities calculated from an ASR system trained on native French speech to evaluate the pronunciation of French learners. They showed that the posterior based measurements provided the highest correlation to human scores. Goodness of pronunciation (GOP), which is the log-posterior probability of aligned phonemes normalized by phoneme duration, was used in \cite{witt2000phone} to detect mispronunciation in non-native speech. The log posteriors in the GOP are also derived from an ASR acoustic model trained on native speech. New advancement in Deep Neural Networks (DNN)-based acoustic models boost the performance of ASR systems, and at the same time these acoustic models have been applied to pronunciation assessment or mispronunciation detection \cite{hu2015improved}\cite{tao2016exploring}. Although some studies also train another ASR system with accented speech in order to generate better recognition/alignment results (thus better fluency and rhythm based features), those measurements related to pronunciation are still extracted from acoustic models trained on native speech \cite{tao2016exploring}\cite{qian2017bidirectional}.

Inspired by the SLM, in this paper we propose to use pronunciation measurements derived from both L1 and L2 acoustic models. We anticipate those features extracted from the L1 acoustic model can provide extra information about the speaker's L2 pronunciation quality. Specifically, two sets of phoneme-level acoustic model confidence scores are implemented: the first is based on the L2 acoustic model (as in \cite{witt2000phone}); the second one utilizes the forced alignment information derived from the L2 acoustic model and extracts a confidence score of the most likely phoneme from the L1 acoustic model phone sets. These confidence scores represent a projection of the speaker's acoustics on the L1 acoustic model and the L2 acoustic model. One set of features estimates the distance of a phoneme pronounced by non-native speaker from native-like pronunciation and the other one estimates the distance of a phoneme pronounced nonnatively in L2 to the closest phoneme in the speaker's L1. Furthermore, we designed an utterance-level feature extraction scheme based on phoneme-level measurements, which can be concatenated and used as an input to a statistical model to predict a pronunciation score. Both implementations are open-sourced.

To the authors' knowledge, there is only one study that uses both L1 and L2 acoustic models to extract measurements for automatic pronunciation evaluation. The authors in \cite{moustroufas2007automatic} used utterance-level confidence scores extracted from both L1 and L2 acoustic models, calculated frame-wise and averaged over the utterance. However, our proposed system has an important difference: our confidence scores are calculated on phoneme segments and provide more specific information regarding the accentedness of different phonemic categories. Furthermore, in \cite{moustroufas2007automatic}, the authors assume the human evaluator can speak both L1 and L2 and experiments were conducted on only one L1. In our study, we want to investigate if the L1 acoustic model can help improve prediction even if the human evaluators have no knowledge of the underlying L1; we carry out experiments with 4 L1s, including Mandarin, Spanish, German and French. The target language is always American English.

We evaluate the proposed system on an accented speech dataset based on a subset of the GMU speech accent archive \cite{weinberger2013speech}. Accentedness scores are collected on Amazon Mechanical Turk (AMT) with judgements from 13 human evaluators for each speaker. Utterance-level features are extracted and sent to a linear regression model. Leave-one-speaker-out cross validation is used to measure the consistency between model predictions and human scores. We show that the proposed system has better consistency with human evaluators compared to systems that only use the target language acoustic model.

\section{Datasets and Methods}

\subsection{Datasets and accentedness annotaion}
\label{sec:datasets}

{\noindent \bf Native speech corpus:} To build the target language acoustic model (English for this study), we use the LibriSpeech corpus \cite{panayotov2015librispeech} and the corresponding training scripts\footnote{https://github.com/kaldi-asr/kaldi/tree/master/egs/librispeech/s5} in the Kaldi toolkit \cite{povey2011kaldi}. The final acoustic model is a triphone model trained with Gaussian Mixture Model-Hidden Markov Model on 960 hours of speech data. The DNN based model was not used because in our experiments we observed that the DNN acoustic model tended to overestimate the pronunciation score. The feature input is a 39-dimensional second order Mel-Frequency Cepstral Coefficient (MFCC) with utterance-level cepstral mean variance normalization and Linear Discriminant Analysis transformation.

For Mandarin, the publicly accessible AIShell Mandarin Speech corpus (approximately 150 hours training data) \cite{bu2017aishell} and the corresponding Kaldi scripts\footnote{https://github.com/kaldi-asr/kaldi/tree/master/egs/aishell/s5} are used. A pronunciation dictionary is included with the dataset. For the remaining three languages (Spanish, French and German), there are no well organized publicly available data. We use data from the Voxforge project and download the speech corpora for French ($\approx$ 30 hours), German ($\approx$ 50 hours) and Spanish ($\approx$ 50 hours). Kaldi scripts\footnote{https://github.com/kaldi-asr/kaldi/tree/master/egs/voxforge/s5} for the Voxforge. The dictionary for these three languages are from the CMU Sphinx (Download available\footnote{https://sourceforge.net/projects/cmusphinx/files/Acoustic\%20 \\ \hspace*{4mm} and\%20Language\%20Models/}). Feature types and structures of acoustic models for the four languages are the same as those used in the English acoustic model.

{\noindent \bf Non-native speech corpus and accentedness annotation:} The non-native speech corpus used in this study is a subset of the GMU speech accent archive \cite{weinberger2013speech} consisting of speakers whose L1s are the aforementioned four languages and native American English. The speakers are chosen carefully to reduce the accent variability and gender imbalance, and to avoid recordings with high background noise. There are 30 speakers for each language, and each speaker reads the same paragraph in English. This results in a dataset with 150 speech recordings. We recruit 13 human evaluators on AMT to rate the accentedness of the 150 speakers with random order and unknown speakers' L1s. The annotators are all native American English speakers and have no or little experience with the four foreign languages. We use a four point annotation scale: 1 = no accent/negligible accent, 2 = mild accent, 3 = strong accent, and 4 = very strong accent. The average duration of the annotation task is $\approx 45 $ minutes and each annotator received \$1.50 (twice the reward in \cite{weinberger2013speech} on similar listening tasks) for their participation in the study.


\begin{figure}[t]
        \begin{minipage}[t]{0.5\linewidth}
        \centering
            \includegraphics[width=1.5in]{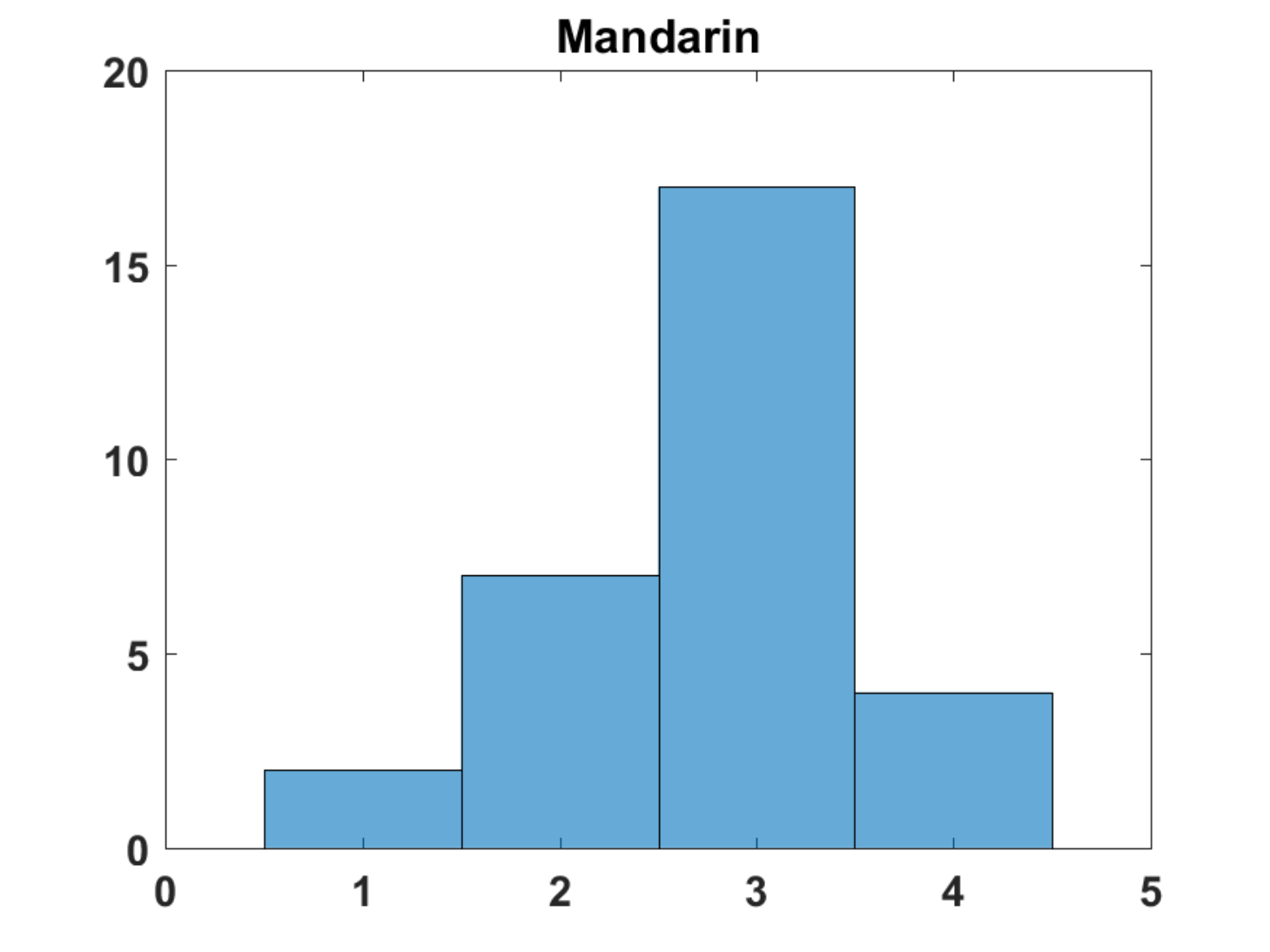}
        \end{minipage}%
        \begin{minipage}[t]{0.5\linewidth}
        \centering
            \includegraphics[width=1.5in]{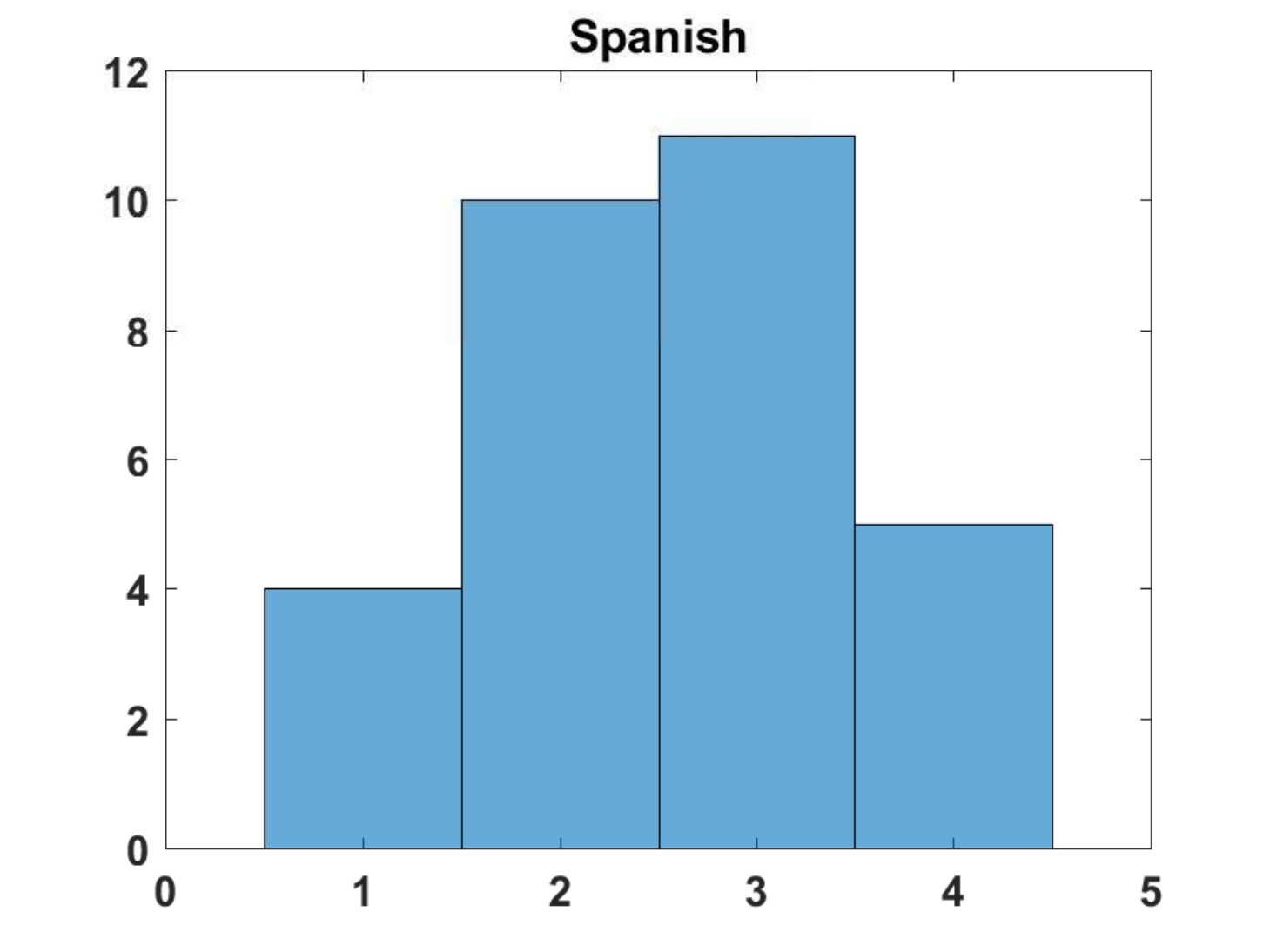}
        \end{minipage}%
        \\
        \begin{minipage}[t]{0.5\linewidth}
        \centering
            \includegraphics[width=1.5in]{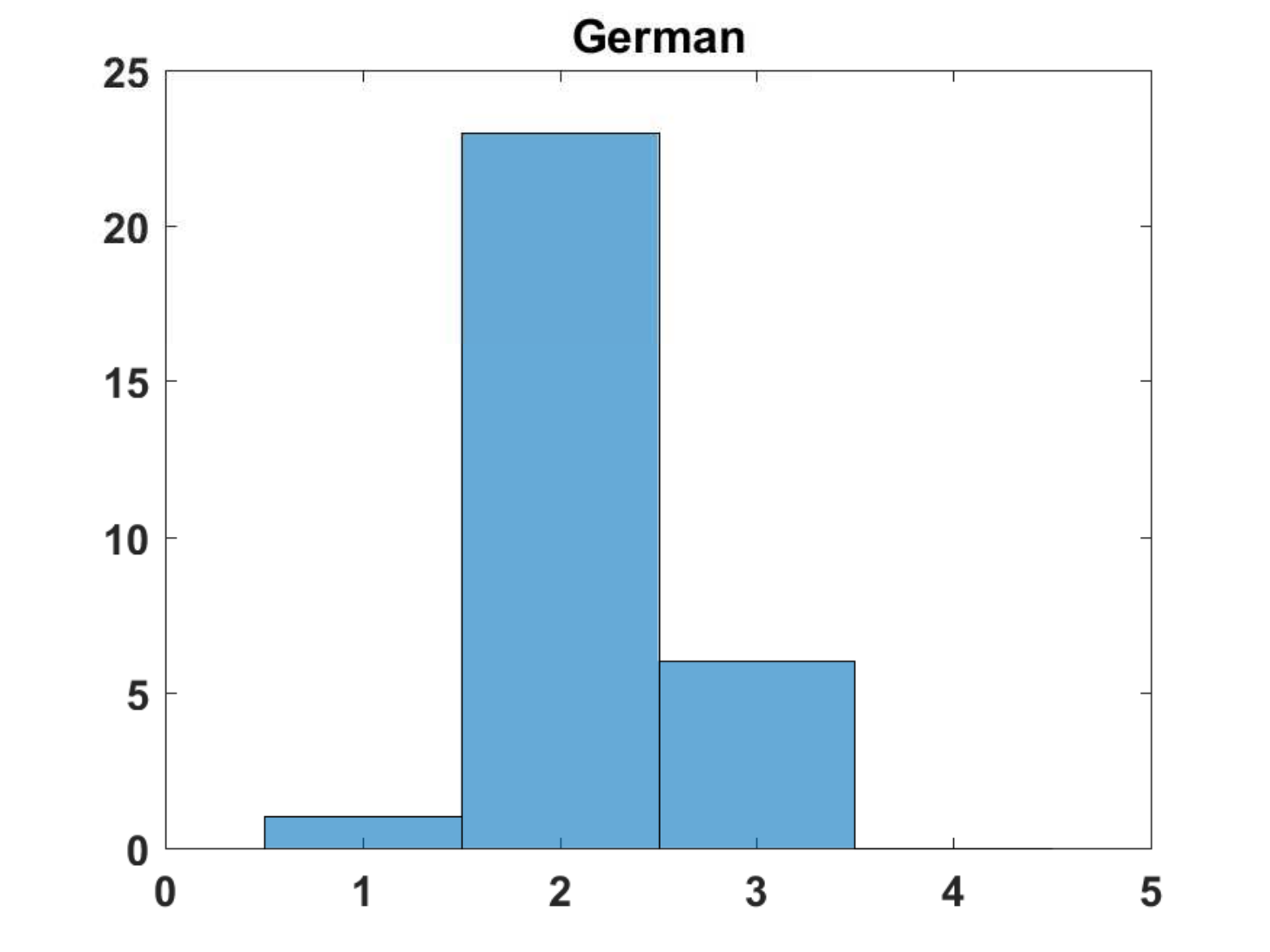}
        \end{minipage}%
        \begin{minipage}[t]{0.5\linewidth}
        \centering
            \includegraphics[width=1.5in]{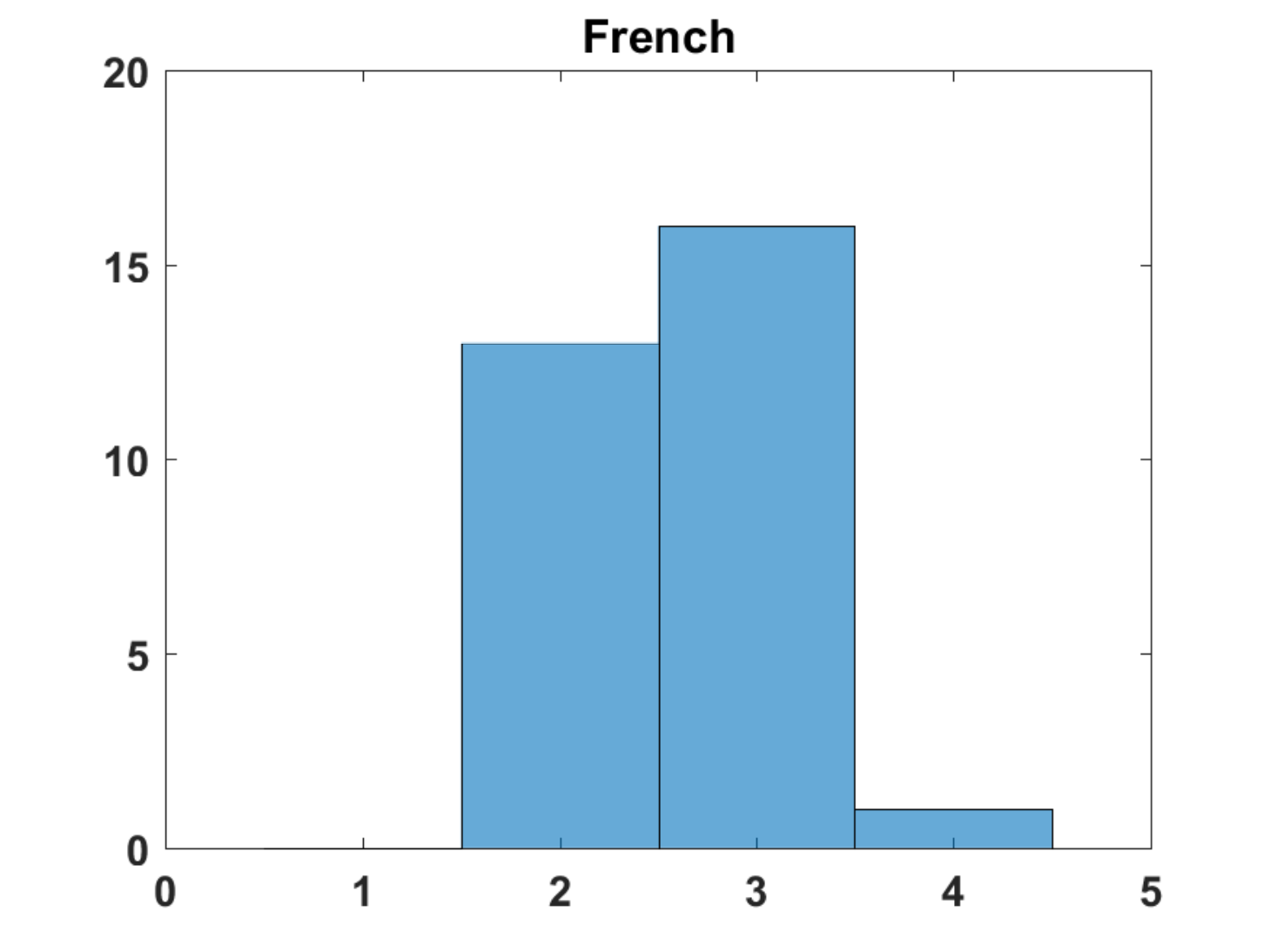}
        \end{minipage}%
        \caption{Histograms of accentedness scores of different L1s.}
        \centering
        \label{fig0}
     \end{figure}

We take the average of all 13 evaluators as the final accentedness rating for each speaker;  other studies have used the average of 10 AMT non-expert annotations in other natural language tasks \cite{snow2008cheap}. The average inter-rater correlation coefficients (calculated as the average of all annotators' correlation with other annotators) is 0.73. In Fig. \ref{fig0}, we show the histograms of the collected ratings across four different foreign languages. Results show that Mandarin speakers have the strongest accent while German speakers have the mildest accent. This is consistent with expectations considering the phonological similarity between German and English as opposed to other 3 languages. For comparison, the average accentedness rating of native English speaker in our dataset is 1.07. The low mean and lack of strongly-accented speakers in the German and French database also means that the variances of the accentedness ratings for these language are relatively low. This poses a challenge in the statistical modeling and will be addressed in section \ref{result}.

\subsection{Feature extraction and system building}

{\noindent \bf Features based on the L2 acoustic model:} Motivated by the work  in \cite{witt2000phone}, we measure the goodness of pronunciation for each phoneme in the accented speech. To do this, the accented speech is first force-aligned at the phoneme-level using the L2 acoustic model to provide the start and end frame indices of each phoneme. We define the pronunciation score ($PS_{\mathrm{L2}}$) of the target phoneme $p$ after alignment as
\begin{equation}
\label{gop}
\begin{aligned}
PS_{\mathrm{L2}}(p) &= \log(P(p|\mathbf{O}^{p}))/\left | \mathbf{O}^{p} \right | \\
      &= \log \left [ \frac{P(\mathbf{O}^{p}|p)P(p)}{\sum_{q\in \mathit{Q}} P(\mathbf{O}^{q}|q)P(q)} \right ] /\left | \mathbf{O}^{p} \right |,
\end{aligned}
\end{equation}
where $\mathbf{O}^{p}$ is the feature matrix of phoneme $p$, $\left |\mathbf{O}^{p}\right |$ is the number of frames of phoneme $p$ after alignment, and $\mathit{Q}$ is the set of all phonemes. If we assume equal priors for all phonemes, we approximate the denominator in Eq. \ref{gop} with max operator,

\begin{equation}
\label{gop2}
PS_{\mathrm{L2}}(p) = \log \left [ \frac{P(\mathbf{O}^{p}|p)}{\max_{q\in \mathit{Q}} P(\mathbf{O}^{q}|q)} \right ] /\left | \mathbf{O}^{p} \right |.
\end{equation}

The conditional likelihood of each phoneme (given the speech frames of the corresponding aligned segment) can be calculated by decoding the sequence of speech features using the L2 acoustic model. It is clear that if the most likely phoneme returned by the acoustic model is the same as the target phoneme $p$, then $PS_{\mathrm{L2}}(p)=0$; otherwise, this value will be negative. The interpretation is that the closer $PS_{\mathrm{L2}}(p)$ is to zero, the closer the pronunciation of phoneme $p$ is to that of native speakers.

{\noindent \bf L1 acoustic model based measurements:} In contrast to the $PS_{\mathrm{L2}}$ score, there does not exist a transcript in L1 for the accented speech to measure pronunciation of the phonemes in L1. We define a new way to calculate the pronunciation score with the L1 acoustic model which quantifies how close the pronunciation of a phoneme in L2 is to a specific phoneme in L1. The forced-alignment calculated with the L2 acoustic model is used here. We first decode the speech frames with the L1 acoustic model and find the state path with the highest likelihood. In the path, the corresponding phonemes of each HMM state are recorded and the phoneme with the highest occurrence is considered as the most likely L1 phoneme for a given speech segment. Then, the pronunciation score is calculated as

\begin{equation}
\label{l1gop}
PS_{\mathrm{L1}}(p) = \left [ \sum_{t \in T_p} \log \frac{ \sum_{s \in S_p}P(o_t|s)}  { \sum_{s \in S}P(o_t|s)} \right ] /\left | T_p \right |,
\end{equation}
where $o_t$ is the feature vector for frame $t$ and $p$ is the phoneme with the highest occurrences in the best decoding path of the current segment. $T_p$ is the set of frames where each frame corresponds to an HMM state of phoneme $p$. $S_p$ is the set of HMM states that belong to phoneme $p$ and $S$ is the set of all HMM states. $PS_{\mathrm{L1}}(p)$ essentially quantifies the confidence of the L1 acoustic model that phoneme $p$ was produced for a speech segment. With eq. \ref{l1gop}, a pronunciation score based on the L1 acoustic model can be calculated for each phoneme segment in the original alignment. The implementations of both sections are available on Github\footnote{https://github.com/tbright17/kaldi-dnn-ali-gop}.

\begin{figure}[t]
        \begin{minipage}[t]{\linewidth}
        \centering
            \includegraphics[width=2.4in]{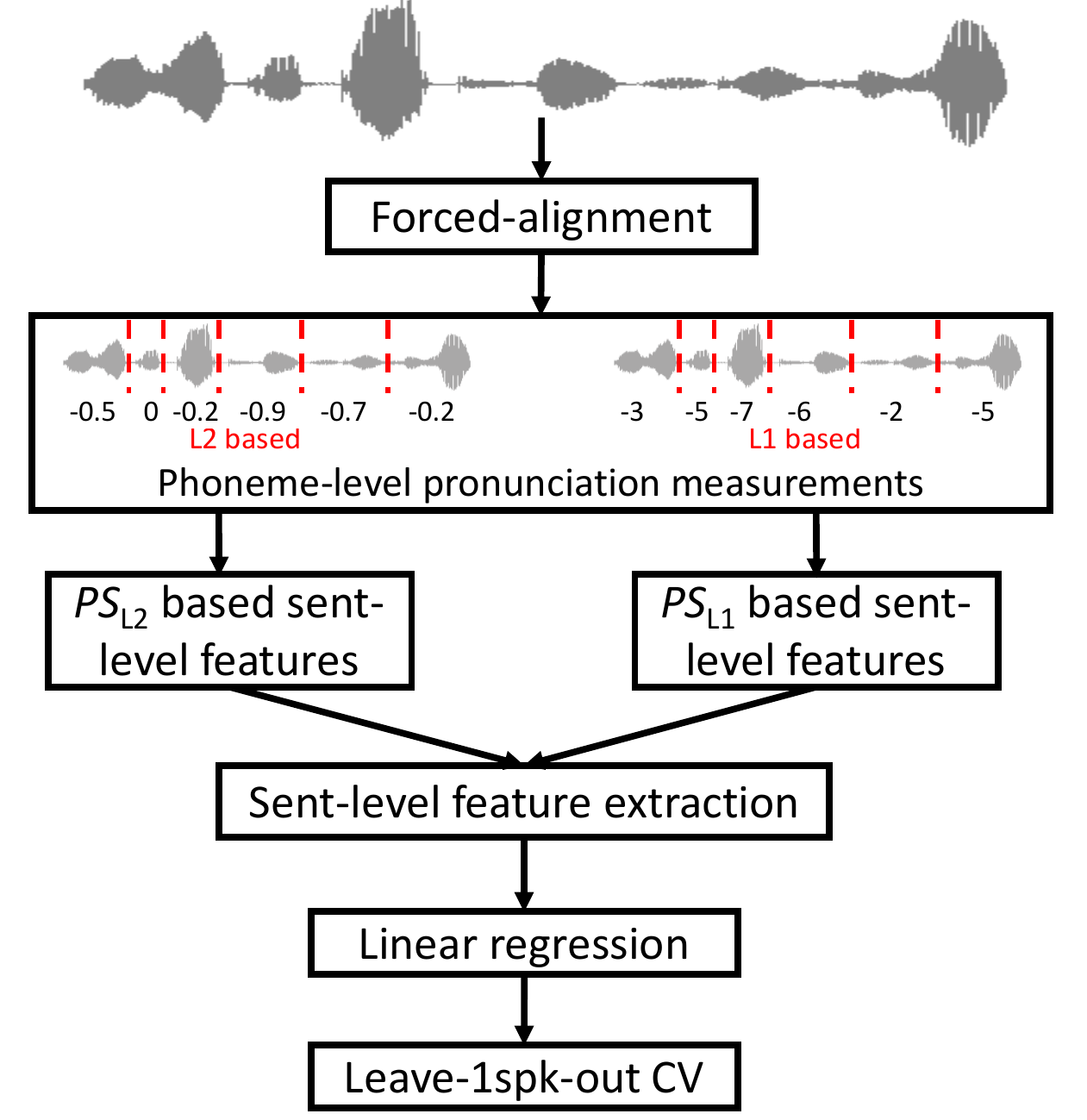}
        \end{minipage}%
        \caption{System diagram.}
        \centering
        \label{fig1}
     \end{figure}

{\noindent \bf Regression-based accentedness prediction:} A diagram of the complete system including forced-alignment, phoneme-level pronunciation score calculation, sentence-level feature extraction and accentedness prediction is shown in Fig. \ref{fig1}. After phoneme-level features $PS_{\mathrm{L2}}(p)$ and $PS_{\mathrm{L1}}(p)$, are extracted, we use a sentence-level feature extraction scheme to convert phoneme-level measurements to a feature vector with a fixed dimension for each utterance. We first combine the pronunciation features for vowels, consonants and syllables and then calculate four statistics for each of these three phonemic categories: for both $PS_{\mathrm{L2}}(p)$ and $PS_{\mathrm{L1}}(p)$, we calculate the minimum, mean, standard deviation and mean-normalized standard deviation (standard deviation divided by mean) of phoneme-level pronunciation measurements of vowels, consonants and syllables in each utterance (implementation available\footnote{https://github.com/tbright17/accent-feat}). This results in a total of 12 utterance-level features for the acoustic model of each language, and a total of 24 utterance-level features combining both pronunciation information from L1 and L2 acoustic models.

To evaluate the predictive ability of this feature set, we build a linear regression model to predict the annotated accentedness from the input feature vector. Since $PS_{\mathrm{L1}}(p)$ is measured with different acoustic models for different languages, we build a different regression model for each L1. We use leave-one-speaker-out cross validation (CV) to estimate the prediction for each test speaker with the remaining speakers used as training data. A simple linear model is used over more complex non-linear models since there are only 30 speakers per language. The system that uses only the 12-dimensional features extracted from only the L2 acoustic model is used as a baseline.

\section{Result analysis}
\label{result}

\begin{figure}[t]
        \begin{minipage}[t]{0.5\linewidth}
        \centering
            \includegraphics[width=1.5in]{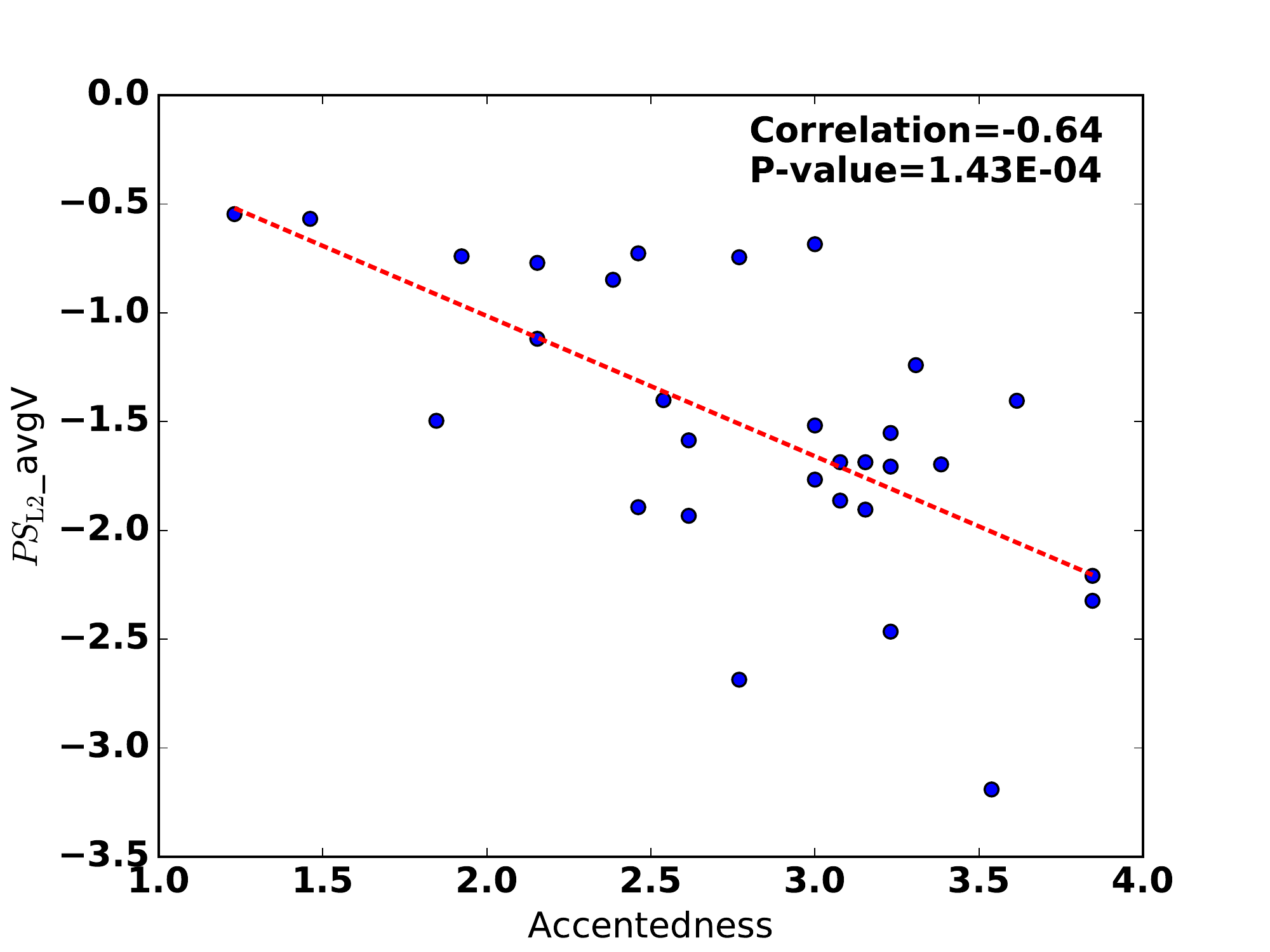}
        \end{minipage}%
        \begin{minipage}[t]{0.5\linewidth}
        \centering
            \includegraphics[width=1.5in]{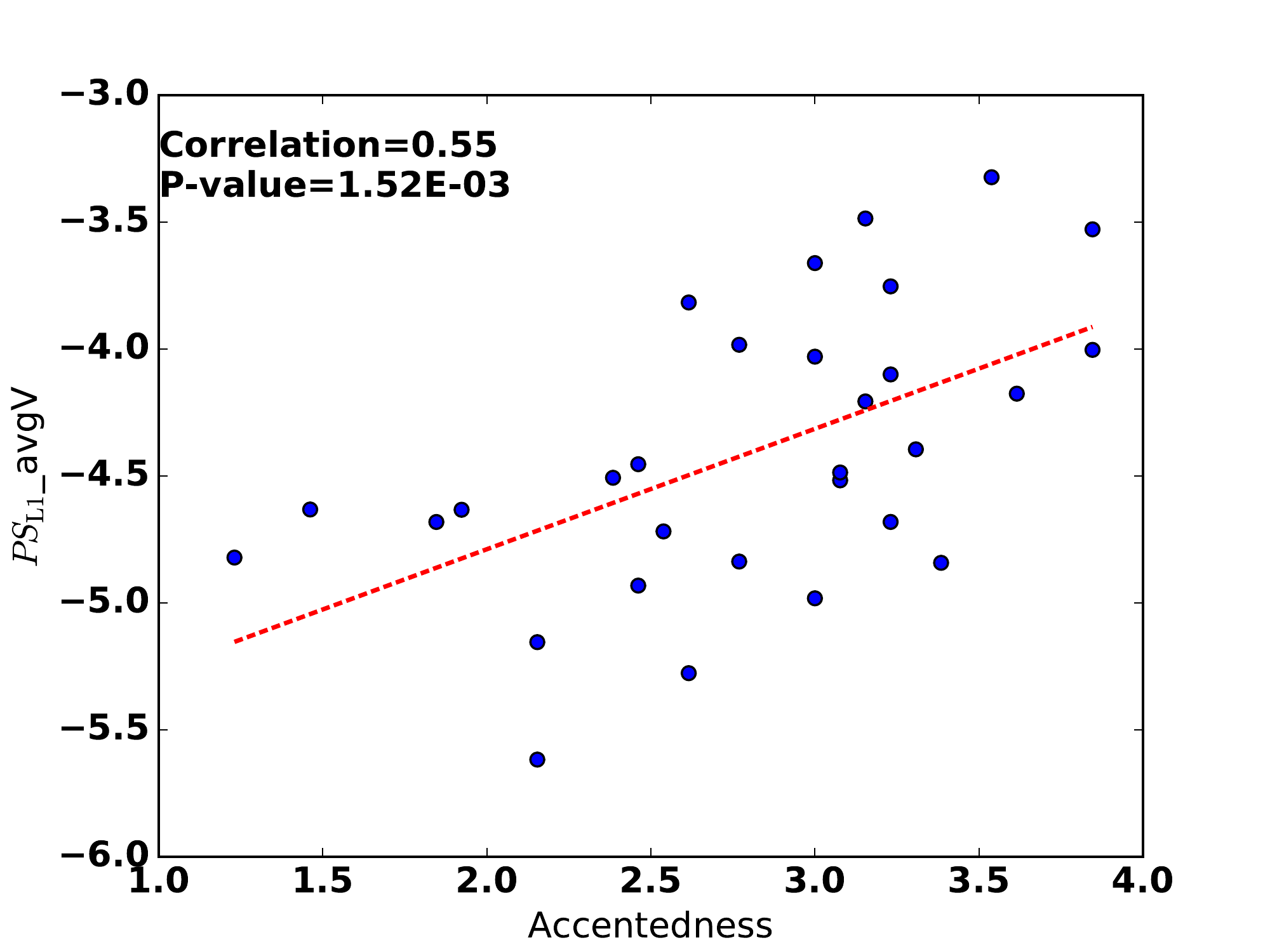}
        \end{minipage}%
        \\
        \begin{minipage}[t]{0.5\linewidth}
        \centering
            \includegraphics[width=1.5in]{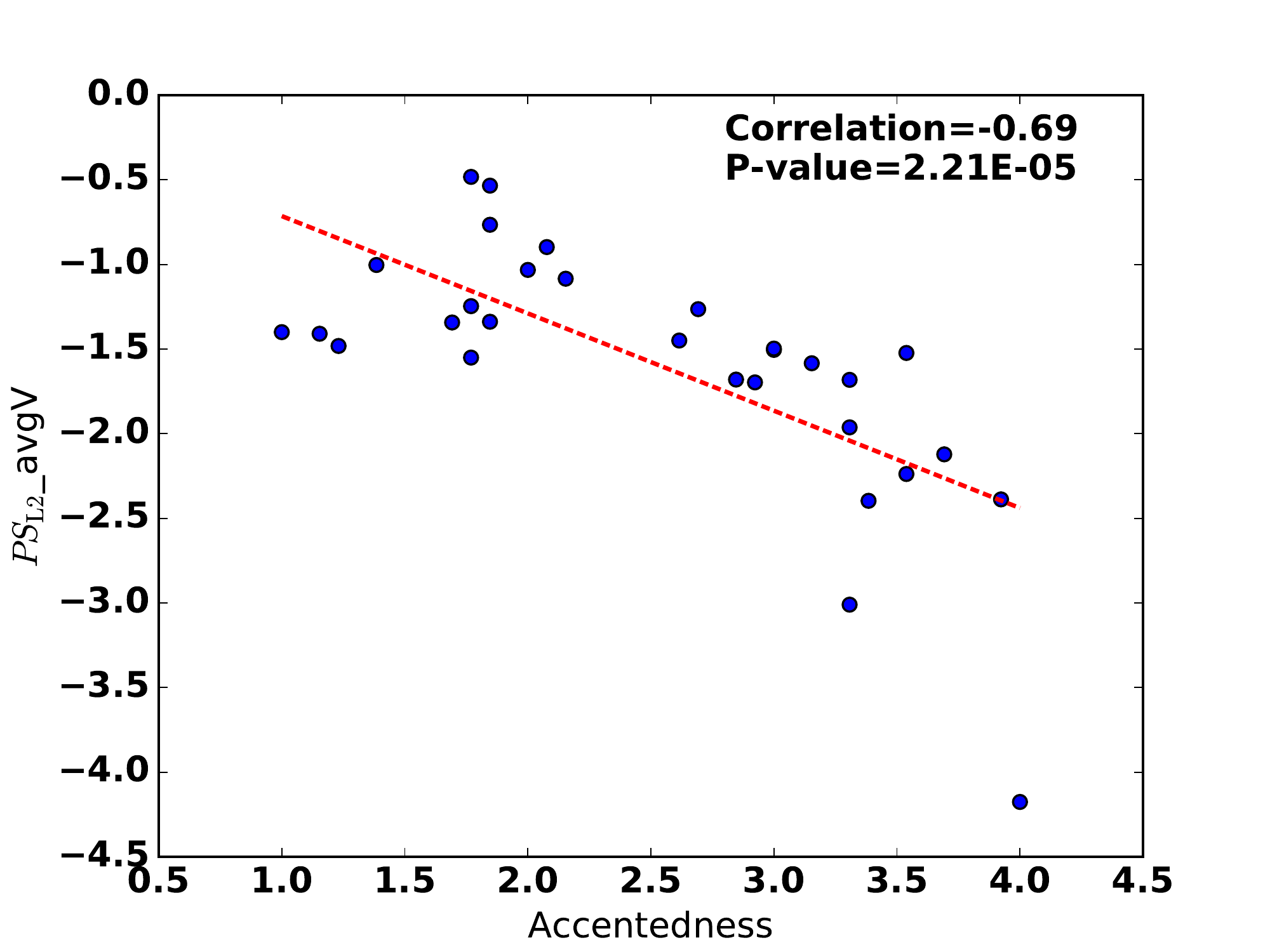}
        \end{minipage}%
        \begin{minipage}[t]{0.5\linewidth}
        \centering
            \includegraphics[width=1.5in]{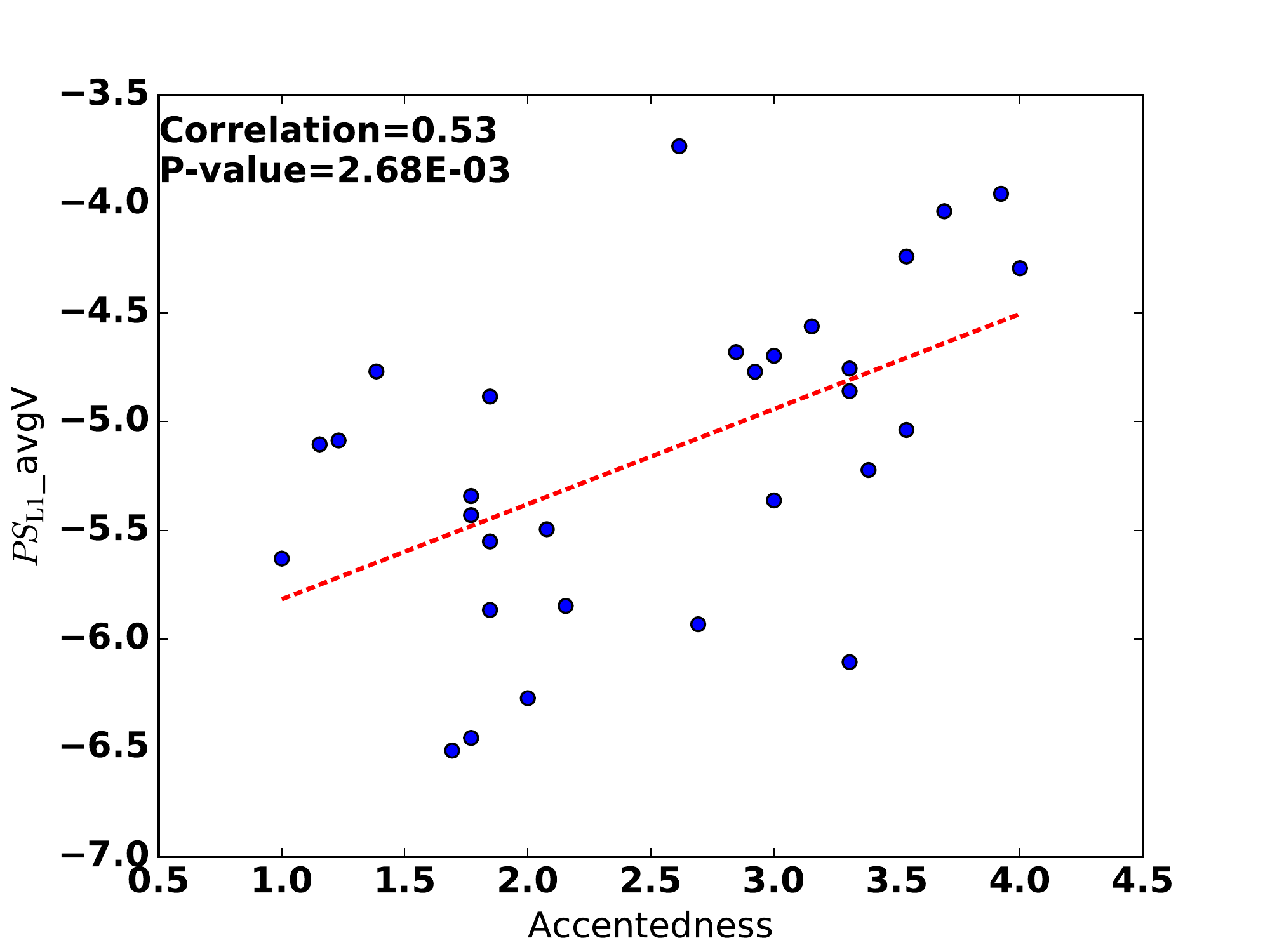}
        \end{minipage}%
        \caption{Scatter plots between accentedness scores and one dimension of features for Mandarin (first row) and Spanish (second row) speakers.}
        \centering
        \label{fig2}
        \vspace{-0.7cm}
     \end{figure}

{\noindent \bf Feature Visualization:} We first illustrate that the extracted pronunciation features provide relevant information regarding the perceived accentedness ratings. In Fig. \ref{fig2} we show four scatter plots relating the accentedness ratings and one of the pronunciation features with Pearson correlation coefficients and statistical significance. The two plots in the first row are for Mandarin speakers. The left plot shows the relationship between the human ratings of accentedness ($X$-axis) and the value of $PS_{\mathrm{L2}}$ averaged over all vowels ($PS_{\mathrm{L2}}$\_avgV on $Y$-axis). The right plot shows the relationship  between the human ratings of accentedness ($X$-axis) and the value of $PS_{\mathrm{L1}}$ averaged over all vowels ($PS_{\mathrm{L1}}$\_avgV on $Y$-axis). The second row shows the same figures for Spanish. It is clear that accentedness and $PS_{\mathrm{L2}}$\_avgV have a negative correlation since larger $PS_{\mathrm{L2}}$\_avgV implies that pronunciation of vowels is closer to native-like pronunciation (and thus a lower accentedness score); accentedness and $PS_{\mathrm{L1}}$\_avgV have a positive correlation since a larger $PS_{\mathrm{L1}}$\_avgV means pronunciation of vowels is closer to L1 pronunciation (and thus higher accentedness score). This provides some confidence that our features exhibit a predictable relationship with accentedness.

{\noindent \bf Accentedness Prediction:} After extracting utterance-level features, each speaker has a feature vector and a corresponding accentedness score (in the range of 1 to 4). For speakers that belong to the same L1 category, a linear regression model with an L2 norm regularizer (or ridge regression) is built with data from 29 speakers used to train the model and the remaining speaker used to evaluate the model. Feature selection based on the univariate linear regression test \cite{saeys2007review} was also used to select the most predictable features. The scikit-learn toolkit was used to implement feature selection and ridge regression \cite{scikit-learn}. To generate the accentedness prediction for all speakers, we perform the evaluation using leave-one-speaker-out CV, which is an unbiased estimate of generalization error \cite{elisseeff2003leave}; this means that a feature selector and a ridge regression model is trained on all combinations of 29 speakers out of 30 speakers and tested on the 1 remaining. For different input features (12-dimensional utterance-level features or 24-dimensional utterance-level features) we tuned hyperparameters for optimal performance.

As mentioned in section \ref{sec:datasets}, the accentedness label distributions for German and French speakers do not span the 1-4 rating scale uniformly. Our initial result revealed that the model performance on German and French speakers was comparatively lower (but there was still improvement over the baseline model). In an attempt to train our model with more uniformly distributed labels, we down-sample the German speakers from 30 to 18 and French speakers from 30 to 22 in an attempt to uniformly sample the labels. For other two languages, there are still 30 speakers in the results. The Pearson correlation coefficient (PCC, higher better) and the mean absolute error (MAE, lower better) are used to measure the relationship between model prediction and human scores.

In table \ref{table2}, we show both the PCCs and MAEs between model predicted accentedness and human annotated accentedness for 4 groups of speakers. We also show the results of German and French speakers before down-sampling in the parentheses. The results for French are comparatively lower; this could be because of the acoustic model was built with a smaller dataset or because of random sampling. In terms of the main purpose of this study, there is a clear improvement when adding $PS_{\mathrm{L1}}$ based features for all 4 L1s. It shows that there is an improvement in model performance consistently and across all languages after adding features from the L1 acoustic model. This is despite the fact that the annotators know little about the acoustic properties of the speakers' L1s.

\section{Discussion}

The results in table \ref{table2} reveal that the improvement in performance varies across different L1s. There are several possible reasons for this including the different modeling quality of the L1s' ASR systems,  the accentedness annotation quality, or the contribution of articulation features to perceived impressions of accentedness for different languages. Another interesting aspect that is worthy of additional investigation is that although there is knowledge transfer from L1 to L2 during L2 acquisition, this influence can vary across different L1s and even different speakers. For example, recent research suggests that there exist some universal effects in L2 learning process that are independent of a speaker's L1 \cite{chang2010first}. Our approach may provide a means of comparing L1-specific and L1-agnostic pronunciation errors to computationally identify some of the universal effects.

We have shown that our proposed feature sets can boost the performance of accentedness prediction. However, there is still room for improvement. First, as mentioned previously, the GMU speech accent archive dataset has a limited number of speakers and small variation of accentedness for some languages. The recording environment also varies by speaker. A cleaner dataset with uniform accentedness ratings is better suited for our application.  Second, the amount and quality of training data for L1 acoustic models can be improved since it is quite limited for some of the languages (Spanish, German and French in this study). More accurate L1 acoustic models may result in an improvement of algorithm performance. Third, it is well known that accentedness is related to both pronunciation and rhythmic features. It is natural to extend the same framework for pronunciation scoring to rhythm features.

\begin{table}[t]
\centering
\caption{PCCs and MAEs between predicted accentedness and human scores for speakers of 4 different L1s.}
\label{table2}
\resizebox{0.9\columnwidth}{!}{%
\begin{tabular}{|c|c|c|c|c|}
\hline
\multirow{2}{*}{} & \multicolumn{2}{c|}{$PS_{\mathrm{L2}}$ only} & \multicolumn{2}{c|}{$PS_{\mathrm{L2}}$ and $PS_{\mathrm{L1}}$} \\ \cline{2-5}
 & PCC & MAE & PCC & MAE \\ \hline
Mandarin & 0.707 & 0.343 & \textbf{0.727} & \textbf{0.329} \\ \hline
Spanish & 0.681 & 0.535 & \textbf{0.730} & \textbf{0.464} \\ \hline
German & \begin{tabular}[c]{@{}c@{}}0.751\\ (0.082)\end{tabular} & \begin{tabular}[c]{@{}c@{}}0.192\\ (0.301)\end{tabular} & \textbf{\begin{tabular}[c]{@{}c@{}}0.833\\ (0.144)\end{tabular}} & \textbf{\begin{tabular}[c]{@{}c@{}}0.163\\ (0.287)\end{tabular}} \\ \hline
French & \begin{tabular}[c]{@{}c@{}}0.371\\ (0.254)\end{tabular} & \begin{tabular}[c]{@{}c@{}}0.373\\ (0.406)\end{tabular} & \textbf{\begin{tabular}[c]{@{}c@{}}0.556\\ (0.411)\end{tabular}} & \textbf{\begin{tabular}[c]{@{}c@{}}0.315\\ (0.370)\end{tabular}} \\ \hline
\end{tabular}}
\end{table}

\section{Conclusions}

In this paper, we used both the L1 and L2 acoustic models to extract features for automatic pronunciation evaluation of accented speech. Two sets of phoneme-level pronunciation measurements are developed to quantify both the deviation of native L2 pronunciation and the similarity with speaker's L1 pronunciation. By combining these two sets of features, we develop a new scheme for extracting sentence-level features to predict human-perceived accentedness scores of accented speech. Experiments on accented speakers from 4 different L1s show that there is an improvement in the model's ability to predict accentedness when pronunciation features from both L1 and L2 are included in the model. Future work includes improving the quality of the L1 models we use in the feature extraction and expanding the model to suprasegmental prosodic features in an attempt to model language rhythm.

\section{Acknowledgements}

The authors gratefully acknowledge the support of this work by an NIH R01 Grant 5R01DC006859-13.


\end{document}